\documentclass[preprint,authoryear,12pt]{elsarticle}

\usepackage{graphics}
\usepackage{amssymb}
\usepackage[%ps2pdf,
a4paper=true,
breaklinks=true,
colorlinks=true,
pdfauthor={E. Atkin et al.},
pdftitle={Secondary cosmic rays in the NUCLEON space experiment}
]{hyperref}
\usepackage{mathtext}
\usepackage[T1,T2A]{fontenc}
\usepackage[utf8]{inputenc}

\journal{Advances in Space Research}

\begin{document}

\begin{frontmatter}

\title{Secondary cosmic rays in the NUCLEON space experiment}

\author[dubna,jinr]{V. Grebenyuk}
\author[sinp]{D. Karmanov}
\author[sinp]{I. Kovalev\corref{cor}}
\cortext[cor]{Corresponding author}
\ead{im.kovalev@physics.msu.ru}
\author[sinp]{I. Kudryashov}
\author[sinp]{A. Kurganov}
\author[sinp]{A. Panov}
\author[sinp]{D. Podorozhny}
\author[jinr,kiev]{A. Tkachenko}
\author[dubna,jinr]{L. Tkachev}
\author[sinp]{A. Turundaevskiy}
\author[sinp]{O. Vasiliev}
\author[sinp]{A. Voronin}

\address[dubna]{“DUBNA” University, Universitetskaya str., 19, Dubna, Moscow region, 141980, Russia}
\address[jinr]{Joint Institute for Nuclear Research, Dubna, Joliot-Curie, 6, Moscow Region, 141980, Russia}
\address[sinp]{Skobeltsyn Institute of Nuclear Physics, Moscow State University, 1(2), Leninskie Gory, GSP-1, Moscow, 119991, Russia}
\address[kiev]{Bogolyubov Institute for Theoretical Physics, 14-b Metrolohichna Str., Kiev, 03143, Ukraine}

\begin{abstract}

The NUCLEON space observatory is a direct cosmic ray spectrometer designed to study cosmic ray nuclei with $Z=1-30$ at energies $10^{12}-10^{15}$~eV. It was launched as an additional payload onboard the Russian Resource-P No. 2 satellite.
In this work B/C, N/O and subFe/Fe ratios are presented. The experiment has worked for half of its expected time, so the data have preliminary status, but they already give clear indications of several astrophysical phenomena, which are briefly discussed in this paper.

\end{abstract}


\end{frontmatter}

\parindent=0.5 cm

\section{Introduction}

Secondary cosmic rays are produced from interaction of primary cosmic ray particles with the interstellar medium (ISM). Their spectra are determined by spectra of primary cosmic rays and features of the ISM \citep[][Ch. 9]{crpp}. Thus measurement of the secondary cosmic ray spectra gives an indirect insight into the ISM of our galaxy.

A lot of experiments have measured and continue to measure spectra of secondary CR: AMS-02 \citep{ams}, ATIC \citep{atic, aticfe}, CREAM \citep{cream}, HEAO \citep{heao-c2, heao-c3fe1, heao-c3fe2, heao-c3fe3}, PAMELA \citep{pamela}, TRACER \citep{tracer1, tracer2}. Their results are in good agreement with expectations and models, but their statistics are limited. 
 
\section{The NUCLEON observatory}

\begin{figure}[th]
\begin{center}
\includegraphics*[width=10cm]{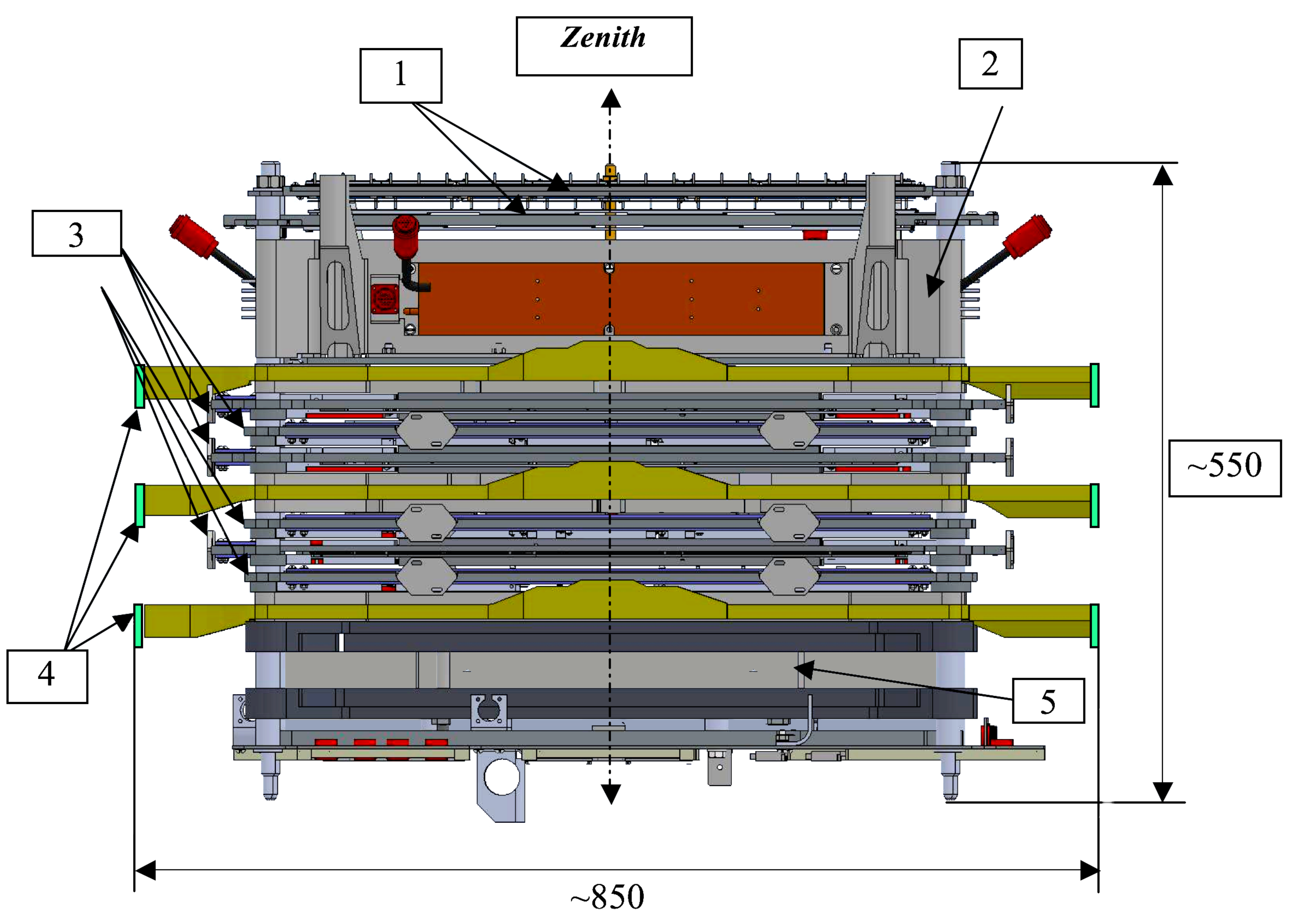}
\end{center}
\caption{The NUCLEON observatory}
\label{nucl}
\end{figure}

The NUCLEON apparatus is a direct measurement cosmic ray spectrometer positioned in the low Earth orbit. The apparatus was designed to measure chemical composition and energy spectra of cosmic rays with charges $Z=1-30$ and energy range of $10^{12}-10^{15}$ eV.

The main feature of the NUCLEON apparatus is a new technique for energy measurement - the kinematic KLEM method. The application of this technique allowed to build a rather small and light device without compromising the geometric factor.

The apparatus is placed on a Russian satellite Resurs-P No. 2 as an additional payload. The satellite is positioned in a low-Earth sun-synchronous orbit with average altitude of 475 km.

The NUCLEON apparatus consists of:

\begin{itemize}
\item The charge measurement system (1), charge resolution $< 0.3$ charge units;
\item The kinematic lightweight energy meter (KLEM) consisting of a carbon target (2) and a silicon tracker (3), energy resolution $\sim 60 \%$;
\item The trigger system (4);
\item The low aperture calorimeter (5), energy resolution $\sim 50 \%$.
\end{itemize}

Details on the KLEM method and its ground tests can be found in these articles: \citet{klem0, klem1, klem2, klem3}.
The analysis method and preliminary analysis results of both the calorimeter and the KLEM can be found in \citet{klemfly}.
Overall design features can be found in \citet{nim}.

\section{Secondary nuclei ratios}

\begin{figure}[th]
\begin{center}
\includegraphics*[width=10cm]{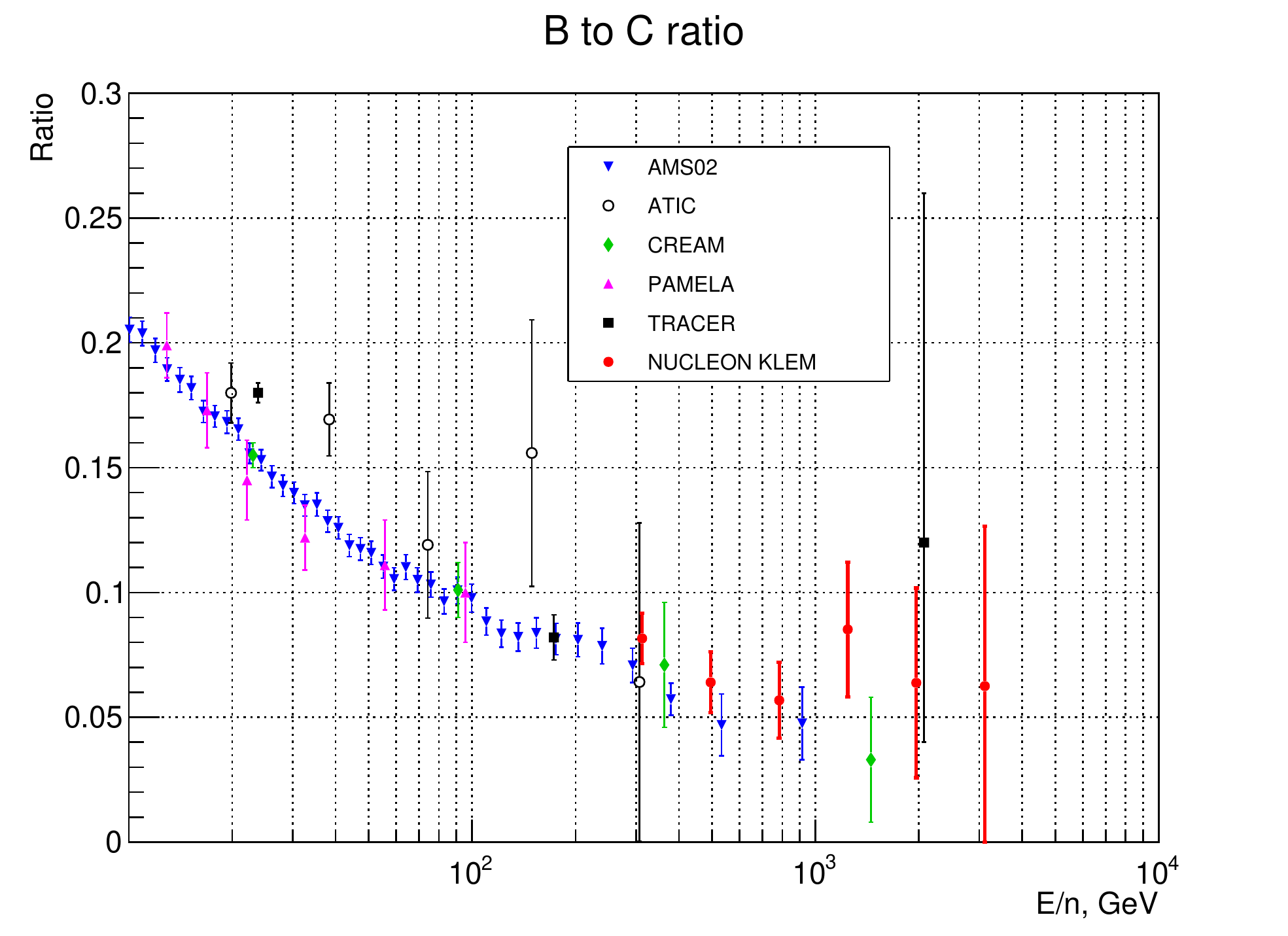}
\end{center}
\caption{This figure shows boron to carbon ratio measured by the NUCLEON experiment as well as other experiments: AMS-02 \citep{ams}, ATIC \citep{atic}, CREAM-I \citep{cream}, PAMELA \citep{pamela}, TRACER \citep{tracer1, tracer2}}
\label{bc}
\end{figure}

Figure \ref{bc} presents the boron to carbon nuclei ratio measured by the NUCLEON experiment compared to other experiments. Only the KLEM system data is presented, because for the IC method the statistics is too low. The B/C ratio is in reasonable agreement with other experiments and has points with energies higher than other experiments.

The B/C ratio is expected to be a decreasing function of energy. All of the previous experiments confirm this expectation except for one point of the TRACER experiment \citep{tracer1, tracer2}, which points to the increase of the ratio but has a significant statistical error. The last 3 points of the NUCLEON experiment also give an indication of the growth of the ratio, but statistical significance of this result is also pretty low. We should point out that the NUCLEON experiment continues data taking, and the statistics should improve.

\begin{figure}[th]
\begin{center}
\includegraphics*[width=10cm]{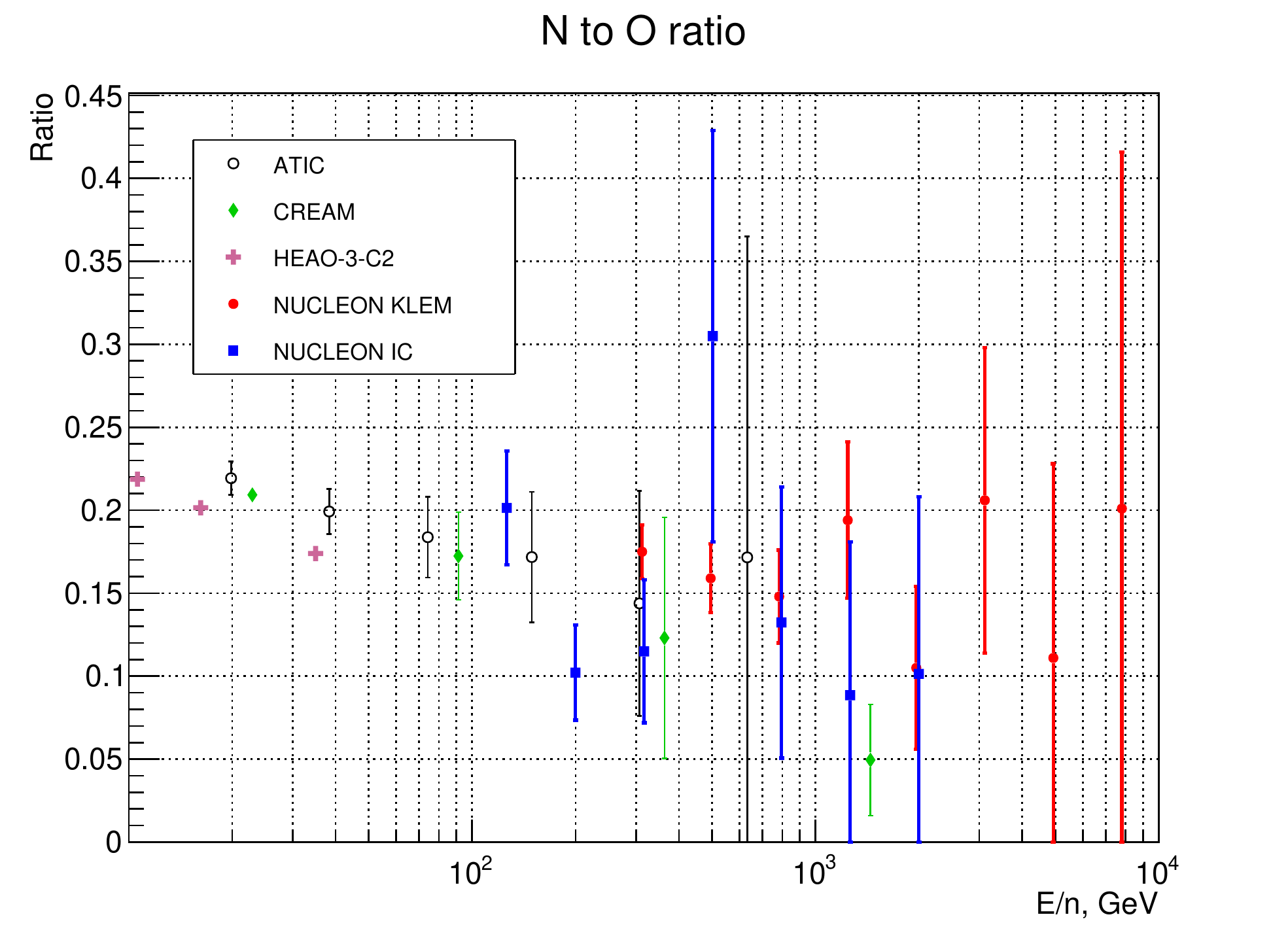}
\end{center}
\caption{This figure shows nitrogen to oxygen ratio measured by the NUCLEON experiment as well as other experiments: ATIC \citep{atic}, CREAM-I \citep{cream}, HEAO-3-C2 \citep{heao-c2}}
\label{no}
\end{figure}

There are similar indications in the nitrogen to oxygen ratio (fig. \ref{no}). Current views show that nitrogen nuclei have partially primary origins, but a significant part of them has secondary nature. Due to that, the N/O ratio should be a decreasing function of energy, like the B/C ratio. The statistical significance of the data is not very high, but there are no clear indication of further decrease of the ratio in the NUCLEON experiment data.

We should note that there are models that predict the increase of the secondary to primary nuclei ratios for high energies \citep[for example, ][]{berezhko}. These models suggest that a significant portion of secondary nuclei is already formed when the cosmic rays are accelerated, inside the termination shock of the supernova remnant. These secondary nuclei can be accelerated in the process of expansion of the supernova shell, and this acceleration can give a bump in ratios for these nuclei at high energies. It is possible that there are indications of such an acceleration mechanism in the NUCLEON experiment data. An increase of the statistics should clarify these indications.

\begin{figure}[th]
\begin{center}
\includegraphics*[width=10cm]{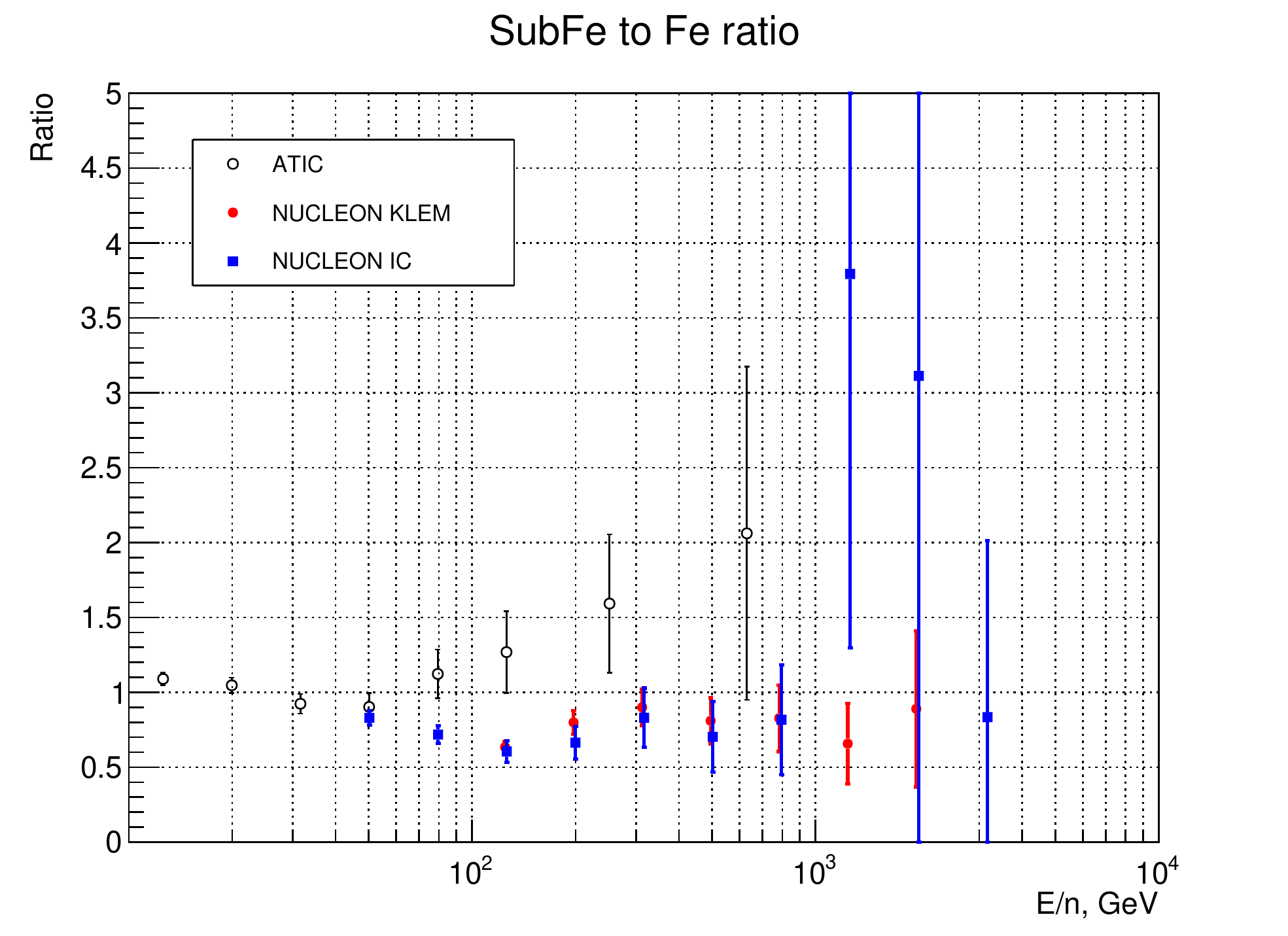}
\end{center}
\caption{This figure shows ratio of nuclei with $Z = 16 - 24$ to iron measured by the NUCLEON experiment and the ATIC experiment \citep{aticfe}}
\label{sfe}
\end{figure}

There is a rather large fraction of secondary nuclei produced by spallation of iron nuclei on ISM, their charges between silicon and iron. Therefore, the ratio the flux of these nuclei to the flux of iron might be expected to be a decreasing function of energy. Ar/Fe and Ca/Fe ratios measured by the HEAO-3-C3 experiment \citep{heao-c3fe1, heao-c3fe2, heao-c3fe3} confirmed these expectations, except that there was an unexpected bump in these ratios for energies over 100 GeV/nucleon. The authors tied it to a possible systematic error in energy measurement procedure. Later, the ratio of the flux of all nuclei with charges from 16 to 24 to the flux of iron was measured in the ATIC experiment \citep{aticfe} and the effect was qualitatively confirmed for this combined ratio.

Figure \ref{sfe} shows the ratio of the spectra of nuclei with $Z=16-24$ to the spectrum of iron nuclei for the NUCLEON experiment and the ATIC experiment \citep{aticfe}. Data from the IC is in qualitative agreement with the indications from the ATIC experiment, although statistical errors are too large. Data from the KLEM method does not show the growth of the ratio compared to the data from the ATIC experiment and to the IC data, but it does not show the decrease either, which is already important.

It is difficult to talk about systematic differences between the results of the calorimeter and the KLEM methods of the NUCLEON experiment, because all the differences occur within the statistical uncertainty. But the issue is expected to be clarified with the increase of the statistics.

\section{Conclusions}

The ratios of secondary nuclei fluxes to primary nuclei fluxes are presented. There are indications of certain features in the presented ratios, although their statistical significance is not high enough. Further analysis of the obtained data and the increase of the statistics should clarify these features.

\section{Acknowledgements}

\end{document}